\documentclass[useAMS,traditabstract]{aa}
\usepackage{eqnarray,amsmath}
\usepackage{natbib,twoopt}
 \bibpunct{(}{)}{;}{a}{}{,}    
\usepackage{graphicx}
\usepackage{txfonts}
\usepackage{ulem}
\usepackage{times}
\begin{document}

\title{Low-redshift lowest-metallicity star-forming galaxies in the SDSS DR14}
\author{Y. I. \ Izotov \inst{1,2}    
\and N. G. \ Guseva \inst{1,2}
\and K. J. \ Fricke \inst{1,3}
\and C. \ Henkel \inst{1,4}
}
\offprints{Y. I. Izotov, yizotov@bitp.kiev.ua}
\institute{          Max-Planck-Institut f\"ur Radioastronomie, Auf dem H\"ugel 
                     69, 53121 Bonn, Germany
\and
                     Bogolyubov Institute for Theoretical Physics,
                     National Academy of Sciences of Ukraine, 
                     14-b Metrolohichna str., Kyiv, 03143, Ukraine
\and 
                     Institut f\"ur Astrophysik, G\"ottingen Universit\"at, 
                     Friedrich-Hund-Platz 1, 37077 G\"ottingen, Germany 
\and
                     Astronomy Department, King Abdulaziz University, 
                     P.O.Box 80203, Jeddah 21589, Saudi Arabia
}
\date{Received \hskip 2cm; Accepted}

\abstract{
We present a sample of low-redshift ($z$~$<$~0.133) candidates for
extremely low-metallicity star-forming
galaxies with oxygen abundances 12~+~logO/H~$<$~7.4
selected from the Data Release 14 (DR14) of the Sloan Digital Sky Survey (SDSS).
Three methods are used to derive their oxygen abundances.
Among these methods two are based on strong  
[O~{\sc ii}]$\lambda$3727$\AA$, [O~{\sc iii}]$\lambda$4959$\AA$, and 
[O~{\sc iii}]$\lambda$5007$\AA$ emission lines, which we call 
strong-line and semi-empirical methods. These were applied for all galaxies. We have developed one 
of these methods, the strong-line method, in this paper. This method is 
specifically focused on the accurate determination of metallicity in extremely
low-metallicity galaxies and may not be used at higher metallicities with
12~+~logO/H~$\ga$~7.5.
The third, the direct
$T_{\rm e}$ method, was applied for galaxies with detected [O~{\sc iii}] 
$\lambda$4363 emission lines. All three methods give consistent abundances 
and can be used in combination or separately for selection of lowest-metallicity
candidates. However, the strong-line method is preferable for
spectra with a poorly detected or undetected [O~{\sc iii}]$\lambda$4363
emission line. In total, our list of selected candidates for extremely 
low-metallicity galaxies includes 66 objects.
}
\titlerunning{The lowest-metallicity SDSS DR14 galaxies}
\authorrunning{Y. I. Izotov et al.}
\maketitle

\section {Introduction}

Nearby dwarf star-forming galaxies (SFG) with extremely low metallicities 
are often considered as local counterparts of 
high-redshift galaxies. Their proximity allows us to study them in
much greater detail than the high-redshift galaxies and to establish
useful constraints on the physical conditions of their low-metallicity
interstellar medium (ISM), the origin of chemical elements, and to develop
models of stellar evolution. These studies can be used to analyse the physical 
properties of the primeval galaxies at redshifts $z$ $\sim$ 5 - 10, which are 
thought to be dwarf systems responsible for the reionization
of the Universe \citep{O09,WC09,B15a,R13,Robertson15,K16}.

However, the number of known nearby lowest-metallicity SFGs with oxygen
abundances 12 + logO/H $\la$ 7.35 is very low. Among them are 
SFGs with 12~+~logO/H $\sim$ 7.00. These are J0811$+$4730 with 
12~+~logO/H = 6.98$\pm$0.02 \citep{I18}, A198691 with 12~+~logO/H = 
7.02$\pm$0.03 \citep{H16} and
SBS 0335$-$052W with 12~+~logO/H ranging from 6.86 to 7.22 in different
star-forming regions \citep{I09}. Recently, \citet{An19} reported the range 
6.96 -- 7.14 in different H~{\sc ii} regions of DDO 68.
Additionally, \citet{P16} and \citet{K18} derived 12 + logO/H $\la$ 7.35
in nearly a dozen of galaxies in the Lynx-Cancer and Eridanus voids.

Large spectroscopic surveys, such as the Sloan Digital Sky Survey (SDSS)
containing millions of galaxy spectra in its database, open an opportunity to
considerably increase the sample of the extremely low-metallicity SFGs 
with 12 + logO/H $\la$ 7.35 \citep[e.g. ][]{I12,G17}. 
In particular, the SFG with a record low 
luminosity-weighted metallicity, J0811$+$4730, was found in the SDSS.

The empirical diagnostic 
[O~{\sc iii}]~$\lambda$5007/H$\beta$ - [N~{\sc ii}] $\lambda$6584/H$\alpha$
diagram \citep{BPT81} can successfully be used to preselect the candidates to
extremely low-metallicity SFGs. On this diagram, they occupy the region with 
both very low [O~{\sc iii}] $\lambda$5007/H$\beta$~$\leq$~3 and very low 
[N~{\sc ii}] $\lambda$6584/H$\alpha$ $\leq$ 0.1,
far from the main sequence of low-redshift SFGs \citep{I12}. Similarly,
the empirical R$_{23}$ -- O$_{32}$ diagram can be used, where
R$_{23}$=([O~{\sc ii}]$\lambda$3727+[O~{\sc iii}]$\lambda$4959+[O~{\sc iii}]$\lambda$5007/H$\beta$ and O$_{32}$=[O~{\sc iii}]$\lambda$5007/[O~{\sc ii}]$\lambda$3727 \citep{I18}. 

A variety of methods have been developed and can be used for the abundance 
determination. The most reliable method is the direct method, but
it requires the detection, with good accuracy, of the [O~{\sc iii}]$\lambda$4363
emission line for the electron temperature determination. For SFGs with
weak or undetected [O~{\sc iii}]$\lambda$4363 emission lines only strong-line
methods can be used. These methods are based on combinations of
strong emission-line intensities of various elements, such as O, N, S, and Ar.
However, the problem with strong-line methods is that the
intensities of emission lines depend not only on the metallicity but also
on some other quantities, such as the ionization parameter. Many of these
methods were constructed for application to galaxies in a wide range of
metallicities, typically at oxygen abundances 12~+~logO/H $\ga$ 7.4, and often
on expence of the accuracy in the abundance determination at lowest 
metallicities \citep[e.g. ][]{PT05,PG16,N06,Y07}.
This paper focuses on simple
empirical strong-line methods for extremely low-metallicity objects based
on oxygen lines. Emphasizing systems with abundances much lower than solar,
we aim to increase the accuracy of abundance determinations in
SFGs with undetected or weak [O~{\sc iii}]$\lambda$4363 
emission. The construction of this method is also motivated by the fact
that recently a number of new extremely low-metallicity SFGs have been discovered
for which the oxygen abundances were derived using high-quality spectra
\citep[][ and references therein]{I18} allowing for more reliable calibration
with the direct method at lowest metallicities. The new method is applied in the present paper to search for extremely low-metallicity 
candidates in the SDSS Data Release 14.

The data base used for the selection of the lowest-metallicity SFGs is described
in Section \ref{S2}. Methods of the oxygen
abundance determination are discussed in Section \ref{sec:method}.
The results of the abundance determination and the list of the SFGs with
12+logO/H $<$ 7.4 are presented in Section \ref{sec:XMD}.
The application of diagnostic diagrams for the selection of extremely 
low-metallicity SFGs is considered in Section \ref{diagn}. 
Our main conclusions are summarised in Section \ref{S4}.

\begin{figure*}[t]
\hspace{0.0cm}\includegraphics[angle=-90,width=0.48\linewidth]{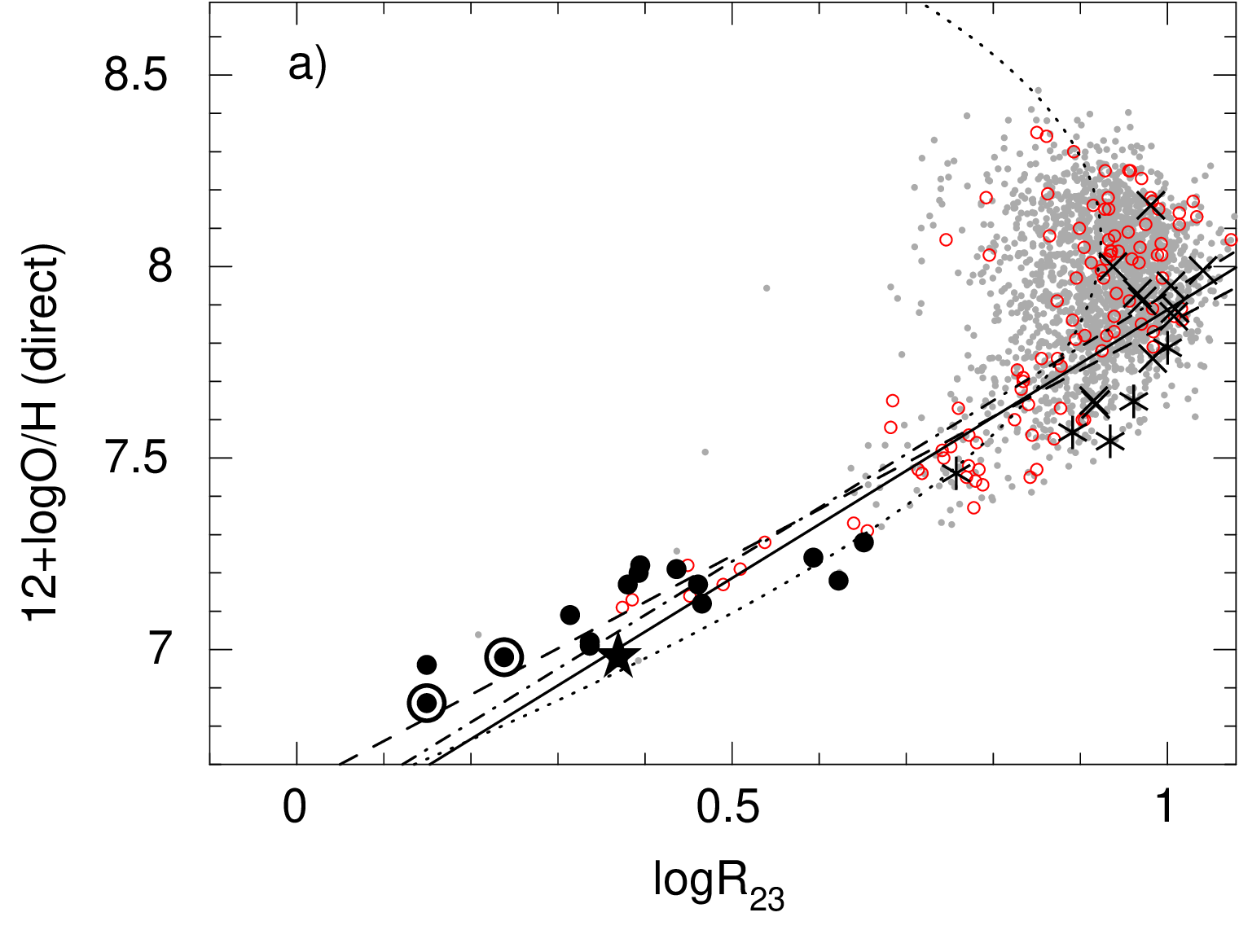}
\hspace{0.2cm}\includegraphics[angle=-90,width=0.48\linewidth]{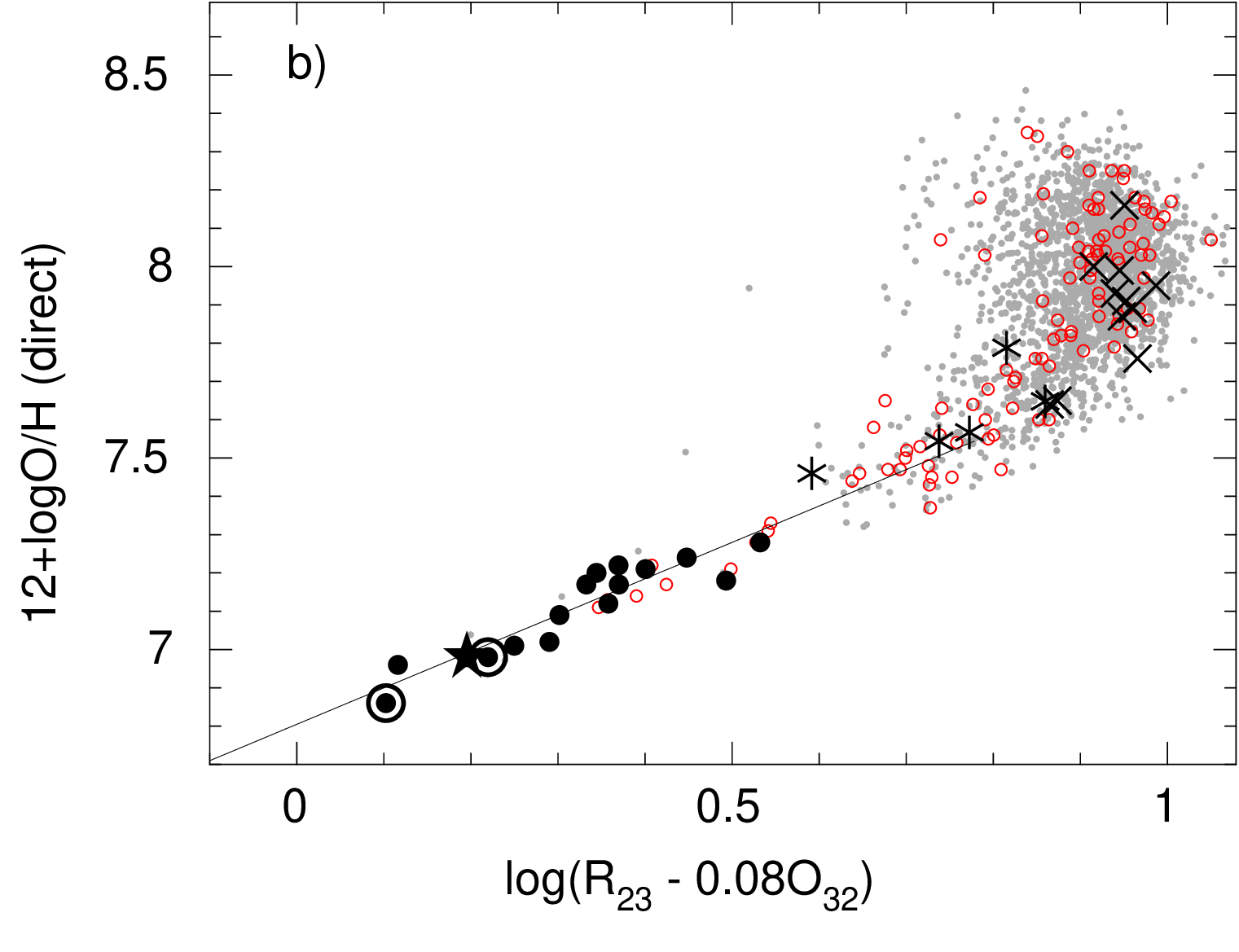}
\caption{{\bf a)} (logR$_{23}$) - (12 + logO/H), and {\bf b)} 
log(R$_{23}$ -- 0.08O$_{32}$) - (12 + logO/H) relations. In both these relations 12 + logO/H is 
derived by the direct method,
R$_{23}$ = ([O~{\sc iii}]$\lambda$4959+$\lambda$5007+[O~{\sc ii}]$\lambda$3727)/H$\beta$ and O$_{32}$ = [O~{\sc iii}]$\lambda$5007/[O~{\sc ii}]$\lambda$3727.
The galaxy J0811$+$4730 with the lowest luminosity-weighted oxygen abundance
\citep{I18} is shown by a filled star, while two knots in SBS~0335$-$052W with
similar or lower oxygen abundances \citep{I09} are represented by encircled 
filled circles. Other lowest-metallicity SFGs from \citet{I18},
the galaxy Little Cub \citep{H17} and the H~{\sc ii} region DDO~68 \#7 
\citep{An19} are shown by filled circles.
SFGs with the highest O$_{32}$ ratios in the range $\sim$ 20 -- 40 \citep{I17} 
are shown by asterisks. Samples of SFGs used for the He abundance 
determination \citep[][ and references therein]{I2014} and Lyman continuum 
leakers \citep[][ and references therein]{I18b} are shown by open circles and
crosses, respectively.
Grey dots represent SDSS SFGs
at $z$ $>$ 0.02 from DR9 and earlier releases and at any redshift from DR10
and later releases, with 
the [O~{\sc iii}] $\lambda$4363\AA\ emission line measured with an accuracy 
better than 25\%. The dashed, dotted, solid, and dash-dotted lines in {\bf a)} 
are the empirical relations by \citet{S89a}, \citet{N06}, \citet{Y07} and
\citet{P00}, respectively.
The solid line in {\bf b)} is the linear most
likelihood fit to the SDSS data with log(R$_{23}$ -- 0.08O$_{32}$) $\leq$ 0.8
(this paper).}
\label{fig1}
\end{figure*}

\section {Data \label{S2}}

 We have searched for extremely low-metallicity galaxies from a sample of 
$\sim$~30000 SFGs selected from the spectroscopic database of the
SDSS DR14 \citep{A18}. This is a continuation of studies which used earlier
SDSS Data Releases and were published by \citet{I12} and \citet{G15,G17}.
Details of SDSS sample selection can be found in \citet{I2014}. 
This sample includes SFGs with equivalent widths
EW(H$\beta$) $\ga$ 10 $\AA$\ of the H$\beta$ emission line in their spectra
indicating an active star formation and the presence of hot massive stars.
The [O~{\sc iii}] $\lambda$4363~$\AA$ emission line is present in spectra
of $\sim$ 18700 out of $\sim$ 30000 SFGs. In $\sim$ 2000 SDSS spectra this 
line is measured with an accuracy better than 25\% allowing for the reliable 
determination of element abundances, most often of oxygen, neon and nitrogen.

  The line fluxes and their errors in each spectrum were measured using 
the {\sc iraf}\footnote{{\sc iraf} is the Image 
Reduction and Analysis Facility distributed by the National Optical Astronomy 
Observatory, which is operated by the Association of Universities for Research 
in Astronomy (AURA) under cooperative agreement with the National Science 
Foundation (NSF).} {\it splot} routine and then corrected for extinction.
 The internal extinction was derived from the Balmer hydrogen emission line  
fluxes after correction for the Milky Way extinction.
The line fluxes were corrected for both reddening \citep{C89}
and underlying hydrogen stellar absorption by the application of an 
iterative procedure \citep{ITL94} and were used for the element abundance
determination. The same SDSS spectra were used for the
determination of some integrated characteristics such as H$\beta$ luminosities,
$L$(H$\beta$), star formation rates SFR, and stellar masses $M_\star$, 
adopting a luminosity distance derived with a cosmological 
calculator \citep[NED,][]{W06}, based on the cosmological 
parameters $H_0$=67.1 km s$^{-1}$Mpc$^{-1}$, $\Omega_\Lambda$=0.682, and
$\Omega_m$=0.318 \citep{P14}.

\section{Methods of oxygen abundance determinations} \label{sec:method}

One of the commonly used methods of the oxygen abundance determination in
low-metallicity SFGs is the direct method based on the determination
of the electron temperature $T_{\rm e}$(O~{\sc iii}) from the
[O~{\sc iii}] $\lambda$4363/($\lambda$4959 + $\lambda$5007) emission-line
ratio and on the relation between $T_{\rm e}$(O~{\sc ii}) and 
$T_{\rm e}$(O~{\sc iii}) obtained by, for example, \citet{I06} from the  
photoionized H~{\sc ii} region models. The knowledge of the
temperatures and the presence of [O~{\sc ii}] and [O~{\sc iii}] emission lines 
in the SDSS spectra allowed us to determine the abundances of these two most
abundant oxygen ions in the H~{\sc ii} regions and thus the total oxygen
abundances for a large sample of SFGs. However, the 
[O~{\sc iii}]$\lambda$4363~$\AA$
emission line is weak in most SDSS spectra of low-metallicity SFGs and is
detected with low signal-to-noise ratio introducing large uncertainties 
in the determination of the electron temperature and oxygen abundance. 

Therefore, for these galaxies, other methods based on strong emission lines
are needed to more reliably derive their metallicities. 
In this paper we have developed a purely empirical strong-line method 
based on the SDSS DR14 sample aiming to apply it for selection of 
the extremely low-metallicity SFG candidates. This method uses the spectroscopic
properties of well-studied
lowest-metallicity galaxies and is calibrated with the direct method.
For the strong-line method the most natural is to use the emission-line fluxes 
of oxygen, the most abundant heavy element. The common approach in the past 
was to use the relation
between the metallicity and the sum of the fluxes of strong oxygen lines 
in the optical range, R$_{23}$~=~([O~{\sc ii}]$\lambda$3727 + 
[O~{\sc iii}]$\lambda$4959 + [O~{\sc iii}]$\lambda$5007)/H$\beta$. 

The problem with this method is that
the relation between the metallicity and R$_{23}$ consists of low- 
(12~+~logO/H~$\la$~8.0) and high-metallicity (12 + logO/H $\ga$ 8.0) branches 
and thus the same value of R$_{23}$ corresponds to two metallicities 
\citep[e.g. ][]{P80,EP84,M91}. Therefore, other constraints separating low- 
and high-metallicity branches are needed to resolve this ambiguity. 
One of the possible
solutions is to use the relation between the metallicities and 
[N~{\sc ii}]$\lambda$6584/H$\alpha$ emission line flux ratios which
monotonically increase with metallicity and thus this relation
is not degenerate \citep[e.g. ][]{V98,PP04}. However, the use of this
relation for the determination of the oxygen abundances
is limited because the [N~{\sc ii}]$\lambda$6584 emission line is
very weak or undetected in low-metallicity galaxies with high-excitation 
H~{\sc ii} regions. We have used this line (or the 
upper limit of its flux) to separate low- and high-metallicity branches.
Adopting [N~{\sc ii}]$\lambda$6584/H$\beta$ $\la$ 0.2
would then select objects located on the low-metallicity branch of
the logR$_{23}$ -- 12~+~logO/H relation.

To construct the logR$_{23}$ -- 12~+~logO/H diagram
we select $\sim$ 2000 SDSS DR14 SFGs in which the [O~{\sc iii}] $\lambda$4363 
emission line is detected with an accuracy better than 25\% (see Sect.~\ref{S2})
allowing for a reliable oxygen abundance determination (grey dots
in Fig.~\ref{fig1}a). Additionally,
all these galaxies show in their spectra [O~{\sc ii}]~$\lambda$3727~$\AA$, 
[O~{\sc iii}]~$\lambda$4959~$\AA$, $\lambda$5007~$\AA$ and have 
[O~{\sc iii}]~$\lambda$4959/H$\beta$~$\ge$~0.2. These data are supplemented
by various samples of SFGs with reliably detected 
[O~{\sc iii}] $\lambda$4363~$\AA$ emission lines in their spectra: a sample of 
extremely low-metallicity SFGs with 12~+~logO/H~$\leq$~7.35 
\citep[filled symbols in 
Fig.~\ref{fig1}a; ][ and references therein]{H17,An19,I18}, a sample
of SFGs with highest O$_{32}$ = 
[O~{\sc iii}]$\lambda$5007/[O~{\sc ii}]$\lambda$3727 ratios of 
$\sim$ 20 -- 40 \citep[asterisks in
Fig.~\ref{fig1}a; ][ and references therein]{I17}, a sample of Lyman continuum 
leaking galaxies with O$_{32}$ ratios in the range $\sim$ 5 -- 28
\citep[crosses in Fig.~\ref{fig1}a; ][ and references therein]{I18b}, and a
sample of SFGs used for the primordial He abundance determination
\citep[open circles in Fig.~\ref{fig1}a; ][ and references therein]{I2014b}.
We note that two encircled filled circles indicate the location
of star-forming regions 3 and 4 in SBS~0335$-$052W \citep{I09} with
undetected [O~{\sc iii}] $\lambda$4363~$\AA$ line. They are shown because of 
their extremely low oxygen abundances, which are among the lowest known and considerably lower 
than the luminosity-weighted oxygen abundance 7.12 of SBS~0335$-$052W 
\citep{IT07}. The electron temperatures and 
the oxygen abundances in these objects are derived by \citet{IT07} using the 
semi-empirical method. However, we have not used them in our subsequent
fitting of relations for the abundance determination.
It is also worth to mention SFG~J0811$+$4730 because of showing the
lowest luminosity-weighted oxygen abundance 12~+~logO/H = 6.98$\pm$0.02
known so far \citep[filled star in Fig.~\ref{fig1}a; ][]{I18}.

\begin{figure*}
\hspace{0.0cm}\includegraphics[angle=-90,width=0.48\linewidth]{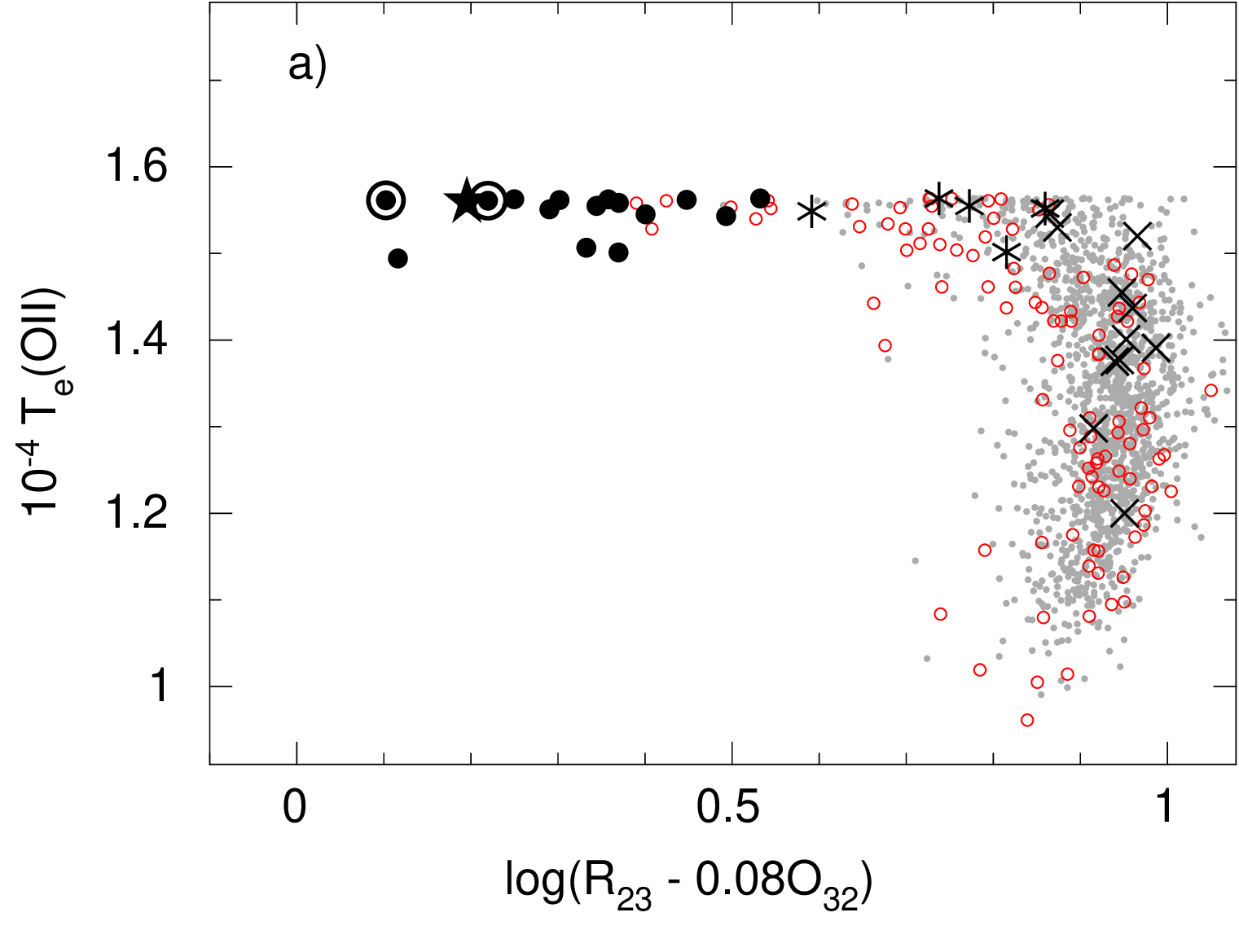}
\hspace{0.3cm}\includegraphics[angle=-90,width=0.48\linewidth]{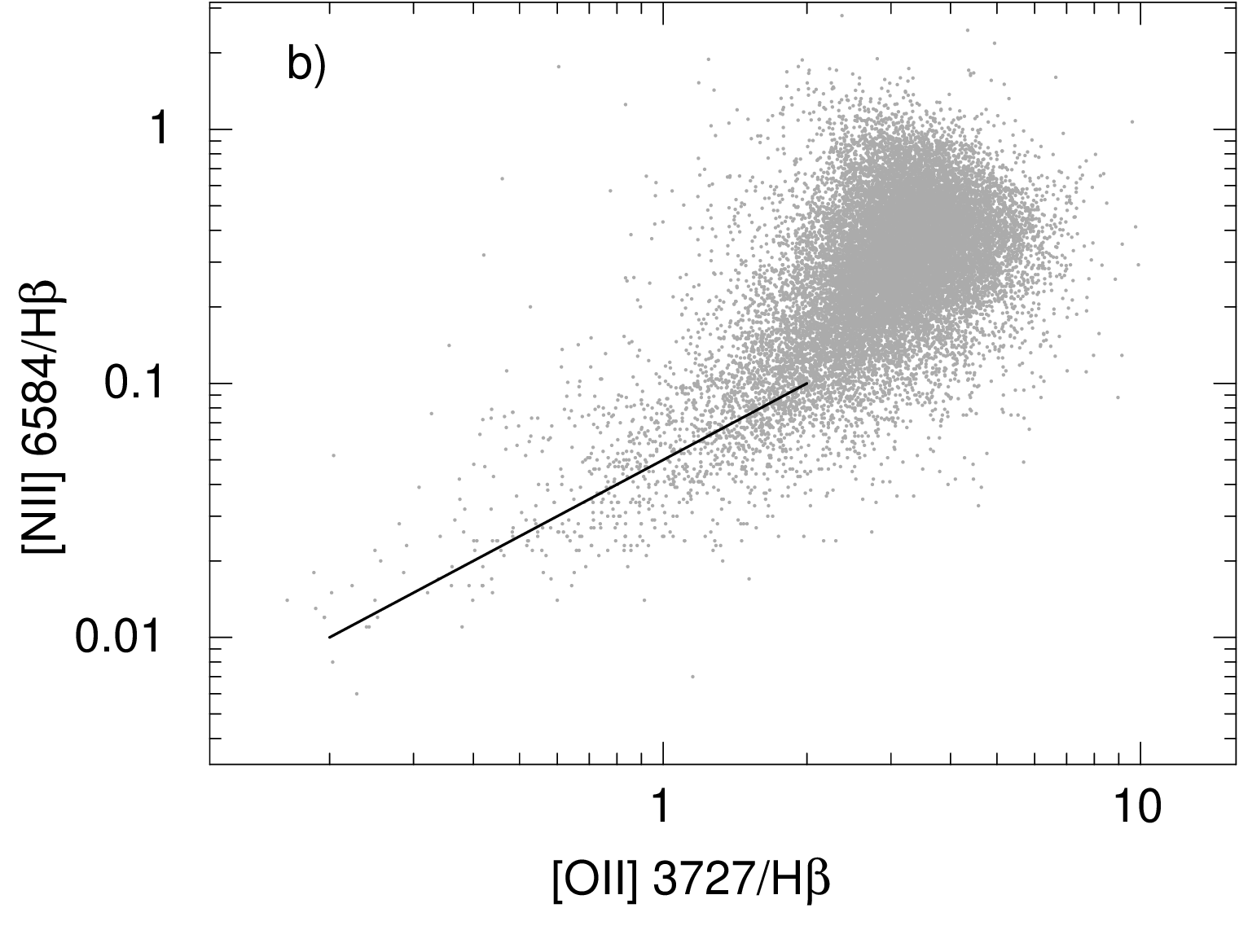}
\caption{{\bf a)} log(R$_{23}$ -- 0.08O$_{32}$) - $T_{\rm e}$(O~{\sc ii}) 
relation for SFGs. {\bf b)} Relation between the extinction-corrected
[O~{\sc ii}]$\lambda$3727/H$\beta$ and [N~{\sc ii}]$\lambda$6584/H$\beta$ 
flux ratios. The solid line corresponds to [O~{\sc ii}]$\lambda$3727 =
20$\times$[N~{\sc ii}]$\lambda$6584. 
Symbols in both panels are as in Fig. \ref{fig1}.}
\label{fig2}
\end{figure*}


It is important that the data include the objects with the highest O$_{32}$
which are indicators of a very high ionization parameter $U$ that is the
measure of the number of ionizing photons per one ion in the
H~{\sc ii} region. The use of these galaxies allows us to eliminate the 
dependence of the calibration relation on $U$.

For the sake of comparison, we show in Fig.~\ref{fig1}a some simple calibration 
relations produced in the past. Three of them, those by \citet{S89a}, 
\citet{P00} and \citet{Y07}, are linear. We note that these relations
reproduce the oxygen abundances reasonably well (within $\sim$ 0.2 dex)
of the extremely low-metallicity SFGs
with log R$_{23}$ $<$ 0.5 shown by filled symbols. On the other hand, SFGs with
highest O$_{32}$ at logR$_{23}$ $>$ 0.5 (asterisks and three SFGs shown by filled
circles) considerably deviate from both the SDSS SFGs and linear relations.

We note the relatively high scatter of filled symbols in the diagram which
we attribute to the ionization parameter varying in a large range. 
To minimize the scatter caused by various ionization parameters we show in
Fig. \ref{fig1}b the relation log(R$_{23}$--0.08O$_{32}$) -- 12~+~logO/H.
The distributions of SFGs 
in Fig. \ref{fig1}b show a much lower scatter than in Fig. \ref{fig1}a.
The data in Fig. \ref{fig1}b can be reproduced by the linear relation
\begin{equation}
12+\log\frac{\rm O}{\rm H} = 0.950\log({\rm R}_{23}-0.08{\rm O}_{32})+6.805, \label{eq:fit}
\end{equation}
shown by a solid line. 


\begin{figure*}
\hspace{0.0cm}\includegraphics[angle=-90,width=0.48\linewidth]{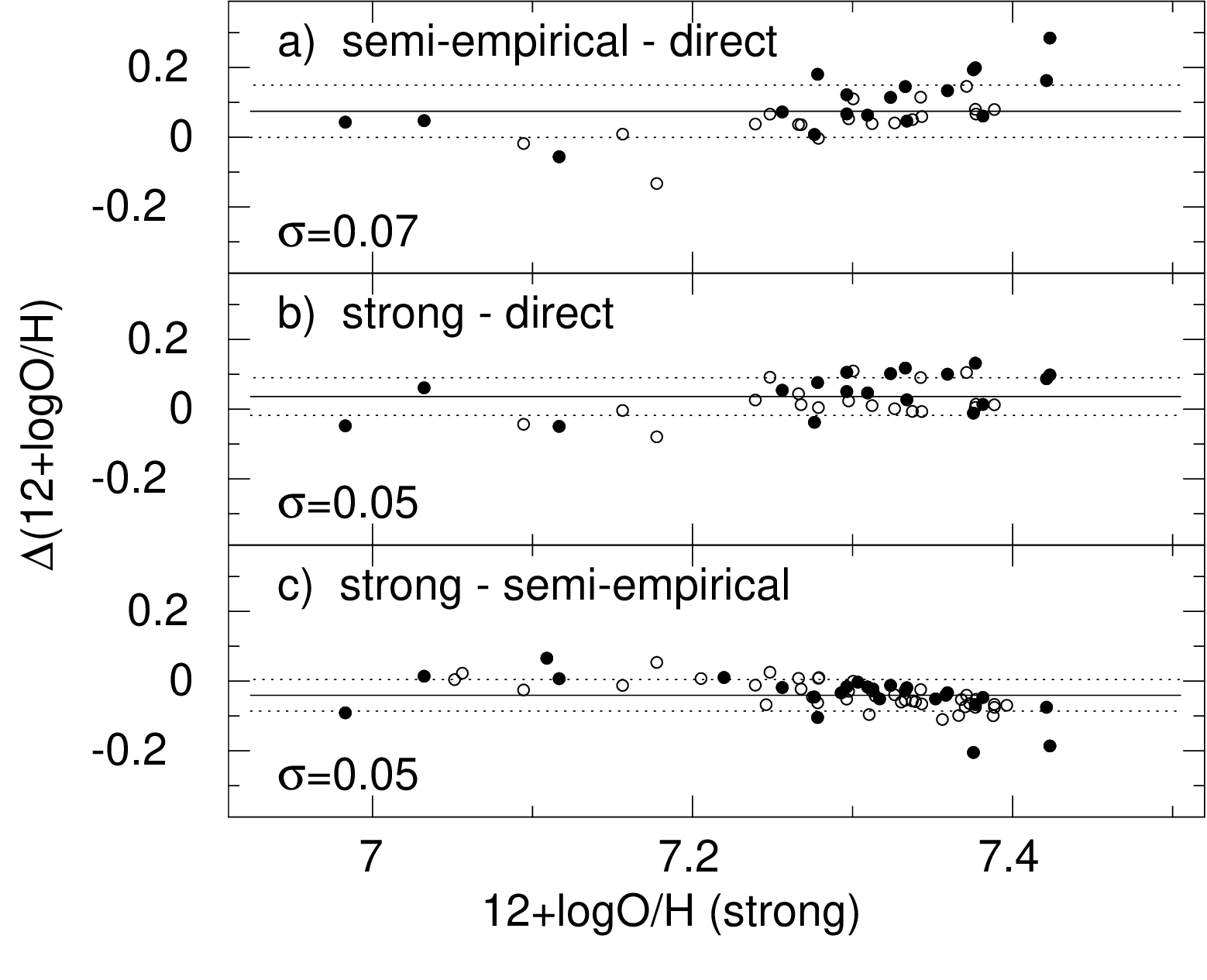}
\hspace{0.5cm}\includegraphics[angle=-90,width=0.48\linewidth]{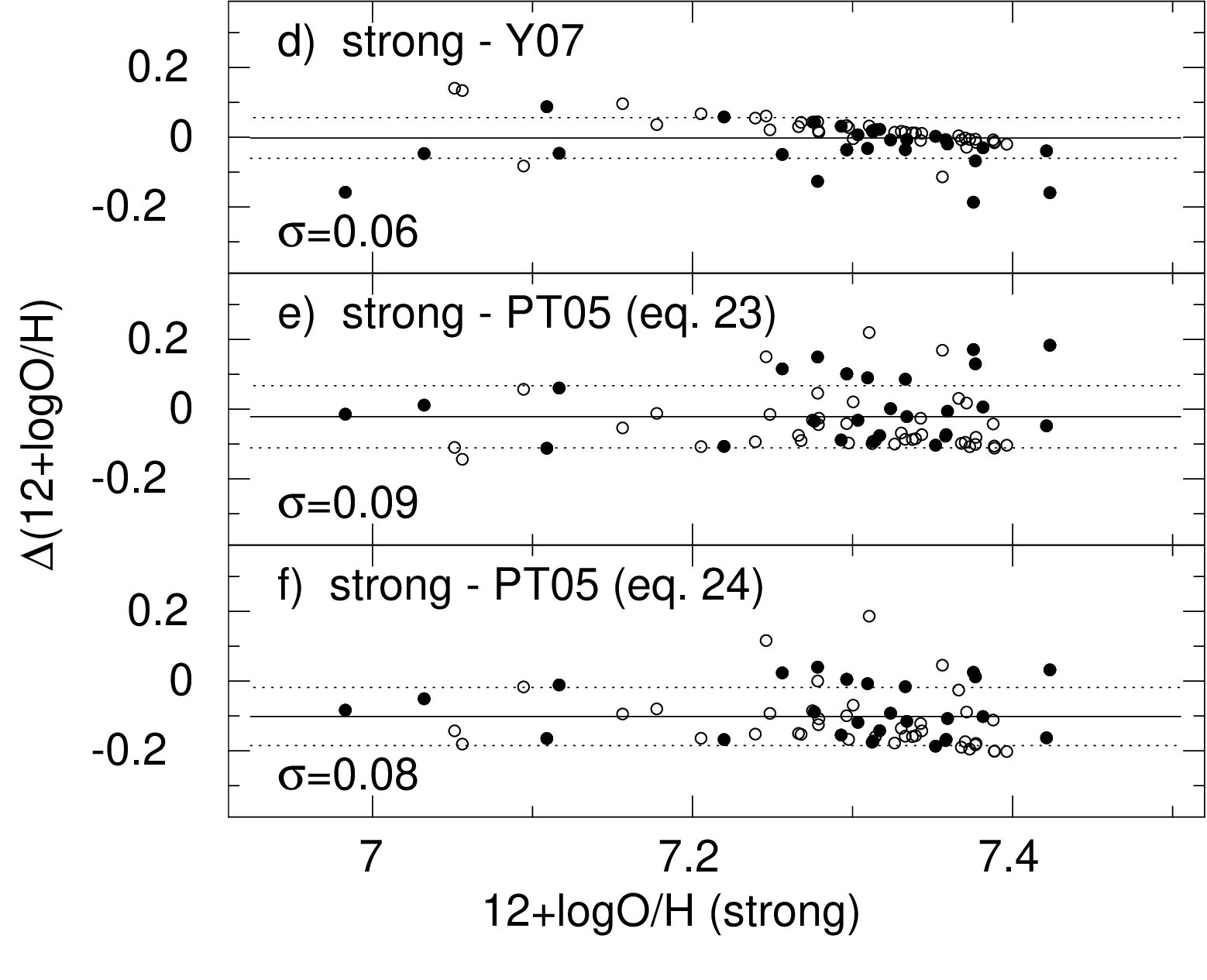}
\caption{{\bf a)} Distribution of differences between oxygen abundances
12 + logO/H derived by the semi-empirical and the direct methods for the sample
from Table \ref{tab2} excluding galaxies with non-detected 
[O~{\sc iii}]~$\lambda$4363$\AA$ emission lines. 
{\bf b)} As in {\bf a)} but for differences
between 12 + logO/H derived by our new strong-line and the direct methods.
{\bf c)} As in {\bf a)} but for differences
between 12 + logO/H derived by the strong-line and the semi-empirical methods
for all SFGs from Table \ref{tab2}. {\bf d)} As in {\bf c)} but for
differences between 12 + logO/H derived by the strong-line method and the 
strong-line method by \citet{Y07}. {\bf e)} As in {\bf c)} but for
differences between 12 + logO/H derived by the strong-line method and the 
P-method \citep[eq. 23 in ][]{PT05}. {\bf f)} As in {\bf c)} but for
differences between 12 + logO/H derived by the strong-line method and the 
P-method \citep[Eq. 24 in ][]{PT05}.
In all panels, galaxies with observed and non-observed 
[O~{\sc ii}]$\lambda$3727 emission line are shown by filled and
open circles, respectively. Solid horizontal lines indicate average abundance 
differences
and dotted horizontal lines indicate $\pm$1$\sigma$ dispersions of the sample 
around the average values.}
\label{fig3}
\end{figure*}



\section{Selection of candidates to the extremely low-metallicity SFGs in SDSS DR14} 
\label{sec:XMD}

We applied Eq.~\ref{eq:fit} for selection of candidates to the extremely 
low-metallicity SFGs with
12~+~logO/H~$<$~7.4. It is seen in Fig.~\ref{fig1}b that these low
metallicities correspond to (R$_{23}$ -- 0.08O$_{32}$) $\la$ 4. We also
adopted an upper limit of [N~{\sc ii}]$\lambda$6584/H$\beta$ $\leq$ 0.2 to 
exclude the contamination of the sample from objects on the upper branch
of the logR$_{23}$ -- 12~+~logO/H relation, and put a low limit
[O~{\sc iii}]$\lambda$4959/H$\beta$ $\ga$ 0.2 to exclude the galaxies which
were not tested in Sect. \ref{sec:method} with the direct method because of
the very weak [O~{\sc iii}]$\lambda$4363~$\AA$ emission line.

The [O~{\sc ii}] $\lambda$3727~$\AA$, [O~{\sc iii}] $\lambda$4959~$\AA$, $\lambda$5007~$\AA$ 
emission line intensities are needed to derive 12 + logO/H by the strong-line
method discussed in Sect. \ref{sec:method}. However, most of SDSS 
lowest-metallicity galaxies are at low redshifts. Therefore, the 
[O~{\sc ii}]~$\lambda$3727~$\AA$ emission line is outside the wavelength range 
of $\sim$ 3800 -- 9200~$\AA$ in
spectra of galaxies with $z$ $\la$ 0.02 selected in DR9 and earlier releases,
precluding the determination of metallicity while spectra of DR10 and
later releases are obtained in a larger wavelength range of
$\sim$ 3600 -- 10300~$\AA$ including the rest-frame wavelength of 
[O~{\sc ii}]~$\lambda$3727~$\AA$ emission line. To avoid this difficulty 
with spectra of DR9 and earlier releases some
prescriptions are needed to estimate the intensity of the
[O~{\sc ii}]~$\lambda$3727~$\AA$
emission line from intensities of other lines. 
In particular, \citet{I12} and \citet{G15,G17} used for that
[O~{\sc ii}]~$\lambda$7320~$\AA$, $\lambda$7330~$\AA$ emission lines. However, 
these lines are relatively weak or undetected
introducing large uncertainties. 
Additionally, for galaxies at redshifts
$z$ $\sim$ 0.02 -- 0.03, their observed wavelengths are close to the wavelength 
$\sim$ 7500~$\AA$ of strong telluric absorption. Keeping in mind these
caveats, we decide to use the 
[O~{\sc ii}]~$\lambda$7320~$\AA$, $\lambda$7330~$\AA$ 
emission lines if detected for the determination of the 
[O~{\sc ii}]~$\lambda$3727~$\AA$ emission line flux from the relation 
\citep{A84,I06}

\begin{equation}
\log I({\rm [O~II]}\lambda 3727) = 0.961 + \frac{8110}{T_{\rm e}({\rm O~II})}+
 \log I({\rm [O~II]}\lambda 7325), \label{I3727}
\end{equation}
where $I$([O~{\sc ii}] $\lambda$7325) = 
$I$([O~{\sc ii}] $\lambda$7320) + $I$([O~{\sc ii}] $\lambda$7330) 
and $T_{\rm e}$(O~{\sc ii}) is the electron
temperature in the O$^+$ zone of the H~{\sc ii} region from where [O~{\sc ii}]
emission lines originate. Thus the electron temperature $T_{\rm e}$(O~{\sc ii}) is
needed to derive $I$([O~{\sc ii}] $\lambda$3727) using Eq. \ref{I3727}.

\begin{figure}
\hspace{0.0cm}\includegraphics[angle=-90,width=0.96\linewidth]{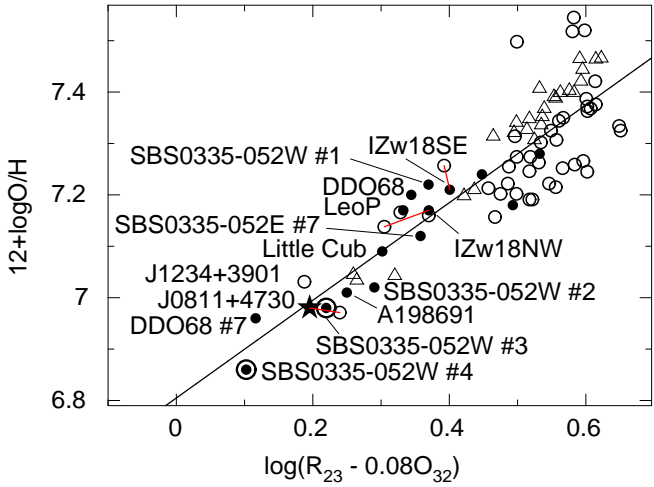}
\caption{Log(R$_{23}$ -- 0.08O$_{32}$) - (12 + logO/H) relation for
SFGs with 12~+~logO/H $\leq$ 7.4 from Table \ref{tab2}. The SFGs from the SDSS 
with 12 +logO/H derived by our newly introduced direct method and by the 
semi-empirical method are shown by open circles and open triangles, respectively.
Other symbols and the solid line are the same as in Fig. \ref{fig1}b. 
Solid lines connect different observations of the same object.
Some extremely low-metallicity SFGs from the literature are labelled.}
\label{fig4}
\end{figure}



\begin{figure}
\hspace{0.0cm}\includegraphics[angle=-90,width=0.96\linewidth]{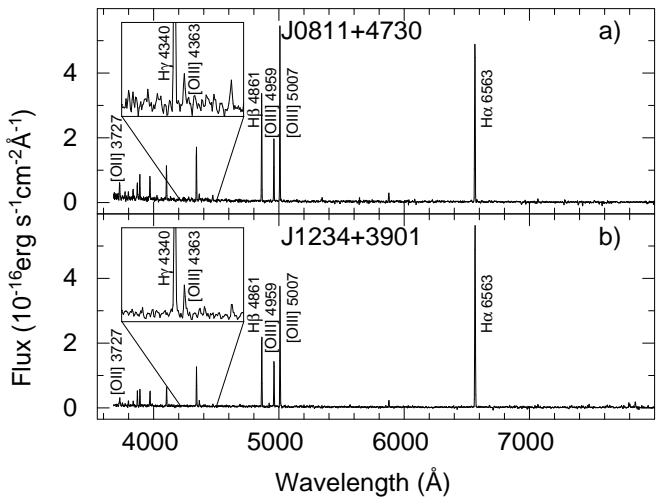}
\caption{SDSS spectra of the two lowest-metallicity galaxies represented
in Table \ref{tab2}. Insets in both panels show expanded segments of the spectra
for a better view of the [O~{\sc iii}] $\lambda$4363~$\AA$ emission lines.}
\label{fig5}
\end{figure}


There is no need to use Eq. \ref{I3727} if [O~{\sc ii}]~$\lambda$3727~$\AA$ is observed.
If it is not observed but [O~{\sc iii}]~$\lambda$4363~$\AA$ is detected, then both
$T_{\rm e}$(O~{\sc iii}) and $T_{\rm e}$(O~{\sc ii}) can be derived. 
The most complicated case for SDSS spectra of 
low-metallicity galaxies is when both [O~{\sc ii}]~$\lambda$3727~$\AA$ 
and [O~{\sc iii}]~$\lambda$4363~$\AA$ emission lines are not detected. Then some 
estimates of the electron temperature $T_{\rm e}$(O~{\sc ii}) are needed to use
the Eq. \ref{I3727}. To do that we consider the relation between 
$T_{\rm e}$(O~{\sc ii}) and log(R$_{23}$ -- 0.08O$_{32}$) for the galaxies with 
spectra where the [O~{\sc ii}]~$\lambda$3727~$\AA$ line is present and the 
[O~{\sc iii}]~$\lambda$4363~$\AA$ line is detected with a relative error better than
25\%. This allowed us to derive both $T_{\rm e}$(O~{\sc ii}) with good accuracy 
and R$_{23}$ -- 0.08O$_{32}$. The relation is shown in Fig. \ref{fig2}a. It is 
seen that $T_{\rm e}$(O~{\sc ii}) at high R$_{23}$ -- 0.08O$_{32}$ varies in a 
wide range, but it is nearly constant with the value of $\sim$ 15500~K at lower
R$_{23}$ -- 0.08O$_{32}$. This constancy is due to the fact that
$T_{\rm e}$(O~{\sc iii}) at low R$_{23}$ -- 0.08O$_{32}$ asymptotically approaches
the value $\sim$ 22000~K. It is also worth to note that {\sc cloudy} 
photoionization
models of hottest H~{\sc ii} regions predict $T_{\rm e}$(O~{\sc iii}) and 
$T_{\rm e}$(O~{\sc ii}) of $\sim$ 22000~K -- 24000~K and $\sim$ 15500~K, 
respectively \citep{F98,F13}.
We adopt the value $T_{\rm e}$(O~{\sc ii}) = 15500~K for the determination of 
$I$([O~{\sc ii}] $\lambda$3727) in the lowest-metallicity
galaxy candidates for which the condition R$_{23}$ -- 0.08O$_{32}$ $<$ 5
is satisfied corresponding to 12~+~logO/H~$\la$~7.5.

We used the extinction-corrected 
[N~{\sc ii}]$\lambda$6584 emission line to estimate the
[O~{\sc ii}]$\lambda$3727 emission-line flux for objects with undetected 
[O~{\sc ii}]$\lambda$7325 emission in their spectra. To find the relation
between [N~{\sc ii}]$\lambda$6584 and [O~{\sc ii}]$\lambda$3727 fluxes we
show in Fig. \ref{fig2}b the diagram [N~{\sc ii}]$\lambda$6584/H$\beta$ --
[O~{\sc ii}]$\lambda$3727/H$\beta$ for SDSS DR14 SFG spectra with detected 
[N~{\sc ii}]$\lambda$6584 and [O~{\sc ii}]$\lambda$3727 lines. We see
that
\begin{equation} 
[{\rm OII}]\lambda3727 = 20\times[{\rm NII}]\lambda6584 \label{eq:NII}
\end{equation}
at low 
[N~{\sc ii}]$\lambda$6584/H$\beta$ $\leq$ 0.1. The determination of the 
[O~{\sc ii}]$\lambda$3727 flux for higher [N~{\sc ii}]$\lambda$6584/H$\beta$ 
is subject to much larger uncertainties and the use of Eq. \ref{eq:NII}
would in general result in overestimation of the [O~{\sc ii}]$\lambda$3727 flux
and correspondingly in overestimation of the oxygen abundance. 
We applied this relation for selection of extremely low-metallicity candidates 
keeping in mind that the oxygen abundance in selected SFGs with 
[N~{\sc ii}]$\lambda$6584/H$\beta$ = 0.1 -- 0.2 might be overestimated.
Nevertheless, additional spectroscopic observations of selected 
galaxies including the [O~{\sc ii}]~$\lambda$3727~$\AA$ emission line are
needed to verify and improve our indirect determination of its flux.

Three methods are used to derive the oxygen abundances. First is the direct
method using prescriptions by \citet{I06}. 
A two-zone photoionised H~{\sc ii} region model was adopted: a high-ionisation 
zone with temperature $T_{\rm e}$(O~{\sc iii}), where [O~{\sc iii}] lines 
originate, and a low-ionisation zone with temperature $T_{\rm e}$(O~{\sc ii}), 
where the [O~{\sc ii}] lines originate. For $T_{\rm e}$(O~{\sc ii}),
the relation between the electron temperatures $T_{\rm e}$(O~{\sc iii}) and
$T_{\rm e}$(O~{\sc ii}) from \citet{I06} is used.
Ionic and total oxygen abundances are derived 
using expressions for oxygen ionic abundances by \citet{I06}. 

The second, semi-empirical method proposed by \citet{IT07} is based on
the determination of the electron temperature from the strong oxygen
emission lines in the galaxies with undetected [O~{\sc iii}]~$\lambda$4363~$\AA$
emission line, while ionic and total oxygen abundances are derived in the way
used by the direct method.

Finally, the third method is the strong-line method (Eq. \ref{eq:fit}) developed
in this paper. This simplest method is purely empirical and does not 
require determination of physical conditions in the H~{\sc ii} region, at 
variance with the two other methods.

We selected 66 extremely low-metallicity SFGs in the entire SDSS 
spectroscopic data base for which at least one of the
three values of 12~+~logO/H derived by each of three methods is less than 7.4.
The list of these SFGs is shown in Table~\ref{tab1} and oxygen abundances 
derived by all three methods
to check their mutual consistency are presented in 
Table~\ref{tab2}. The Tables also include coordinates, redshifts, line 
intensities, equivalent widths of the H$\beta$ emission line, 
and integrated characteristics, such as SDSS absolute magnitudes
$M_g$ in the $g$-band and stellar masses $M_\star$ derived from the fitting
of SDSS spectra. The selected galaxies are located mainly at very low 
redshifts, have faint absolute magnitudes and low stellar masses,
characterising them as dwarf star-forming galaxies. The 
[O~{\sc iii}]~$\lambda$4363~$\AA$ emission line is detected in most selected galaxies
allowing the 12 + logO/H determination by the direct method. 

A notable feature
is that all three methods give consistent results with a dispersion
less than 0.1 dex in most cases (Table~\ref{tab2}). 
In Fig. \ref{fig3} we compare oxygen abundances derived by our strong-line
method with those derived by the direct, semi-empirical, and some other 
strong-line methods in the literature. The agreement between 12+logO/H derived
with the direct method that derived with our newly developed strong-line 
metnod is somewhat better than that derived with the semi-empirical method.
This is more clearly seen in Figs. \ref{fig3}a and \ref{fig3}b, in which we show 
only galaxies with detected [O~{\sc iii}]~$\lambda$4363~$\AA$ emission line.
On average, the oxygen abundances derived by the strong-line and semi-empirical
methods are respectively by $\sim$ 0.04 dex and $\sim$ 0.08 dex higher than 
those derived by the direct method.
In Fig.~\ref{fig3}c we compare oxygen abundances derived by the strong-line
and the semi-empirical methods which on average are consistent within 
$\sim$ 0.05 dex.
The strong-line method developed
in this paper and semi-empirical method are likely more preferable
compared to the direct method for galaxies with weak 
[O~{\sc iii}]~$\lambda$4363~$\AA$ emission, detected
with poor signal-to-noise ratio.

In Fig.~\ref{fig3}d -- \ref{fig3}f we compare our strong-line method with
some previously proposed strong-line methods from the literature. 
\citet{Y07} presented a simple 
relation between 12~+~logO/H and R$_{23}$ which on average gives oxygen 
abundances consistent with our strong-line method (Fig.~\ref{fig3}d). 
However, at lowest 12~+~logO/H $\la$ 7.1, the relation by \citet{Y07} 
predicts oxygen abundances by as much as $\sim$ 0.2 higher than those
obtained with our strong-line method. \citet{PT05} develop their 
strong-line method by introducing the parameter 
P = [O~{\sc iii}]$\lambda$5007/([O~{\sc ii}]$\lambda$3727 + 
[O~{\sc iii}]$\lambda$5007) which takes into account the dependence on
the ionization parameter. They present two relations for their method which
we compare in Figs.~\ref{fig3}e and \ref{fig3}f with our strong-line method.
Two features of this comparison are worth to note: 1) a larger dispersion of
objects compared to that in Fig.~\ref{fig3}c and 2) a systematic offset of
average values of $\Delta$(12~+~logO/H) from the zero value indicating that
the two modifications of the P-method give systematically higher 12~+~logO/H,
by $\sim$ 0.02 (Fig.~\ref{fig3}e) and by $\sim$ 0.10 (Fig.~\ref{fig3}f).
Finally, we note that there is no difference between the galaxies 
with observed and non-observed [O~{\sc ii}] $\lambda$3727 emission line 
in all panels of Fig.~\ref{fig3} (filled and open circles, respectively)
implying reliability of calculated intensities of this line.
From the above comparison we conclude that our simple strong-line method
is likely the most reliable for the oxygen abundance determination of extremely
low-metallicity galaxies.

\begin{figure*}
\hspace{0.0cm}\includegraphics[angle=-90,width=0.48\linewidth]{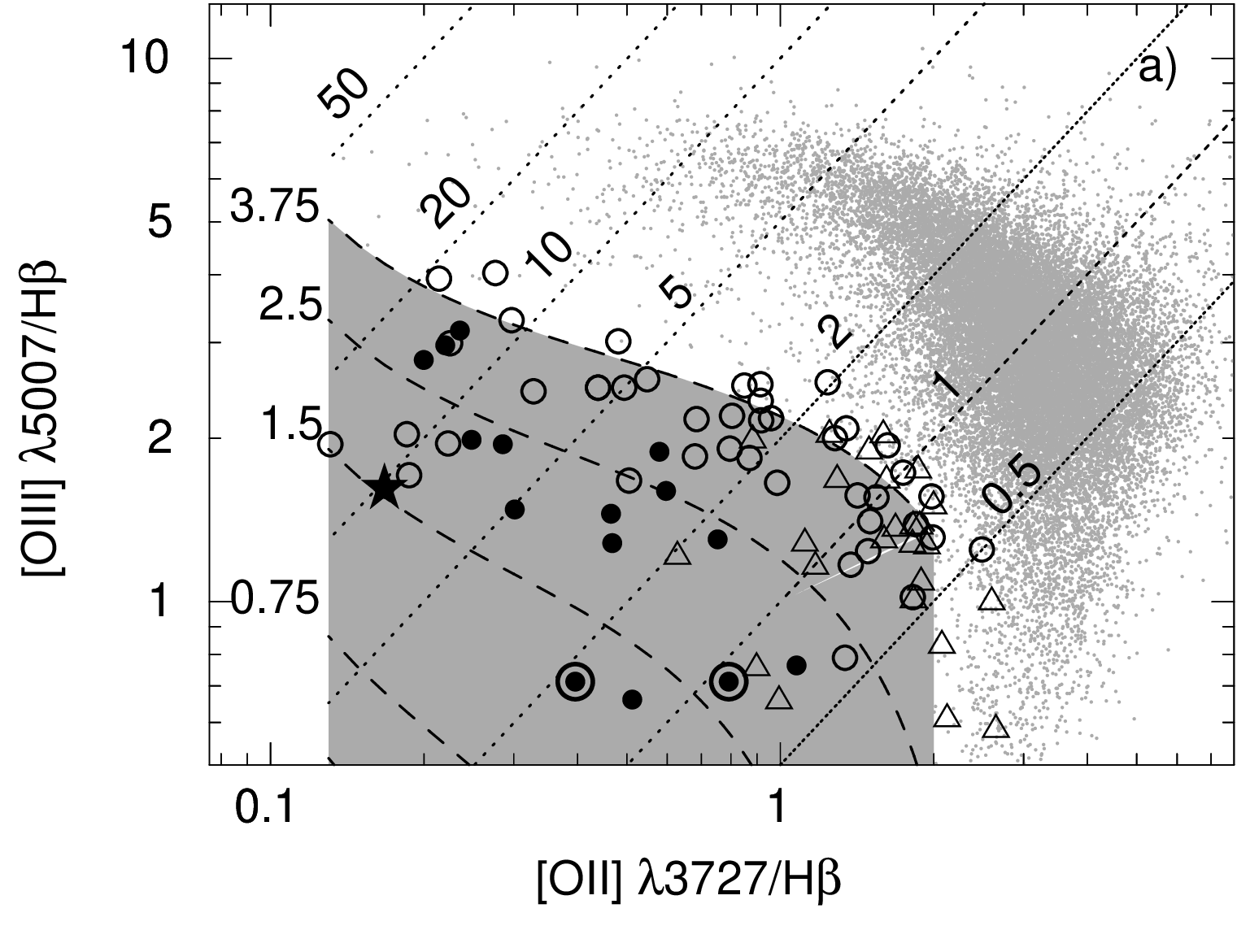}
\hspace{0.2cm}\includegraphics[angle=-90,width=0.48\linewidth]{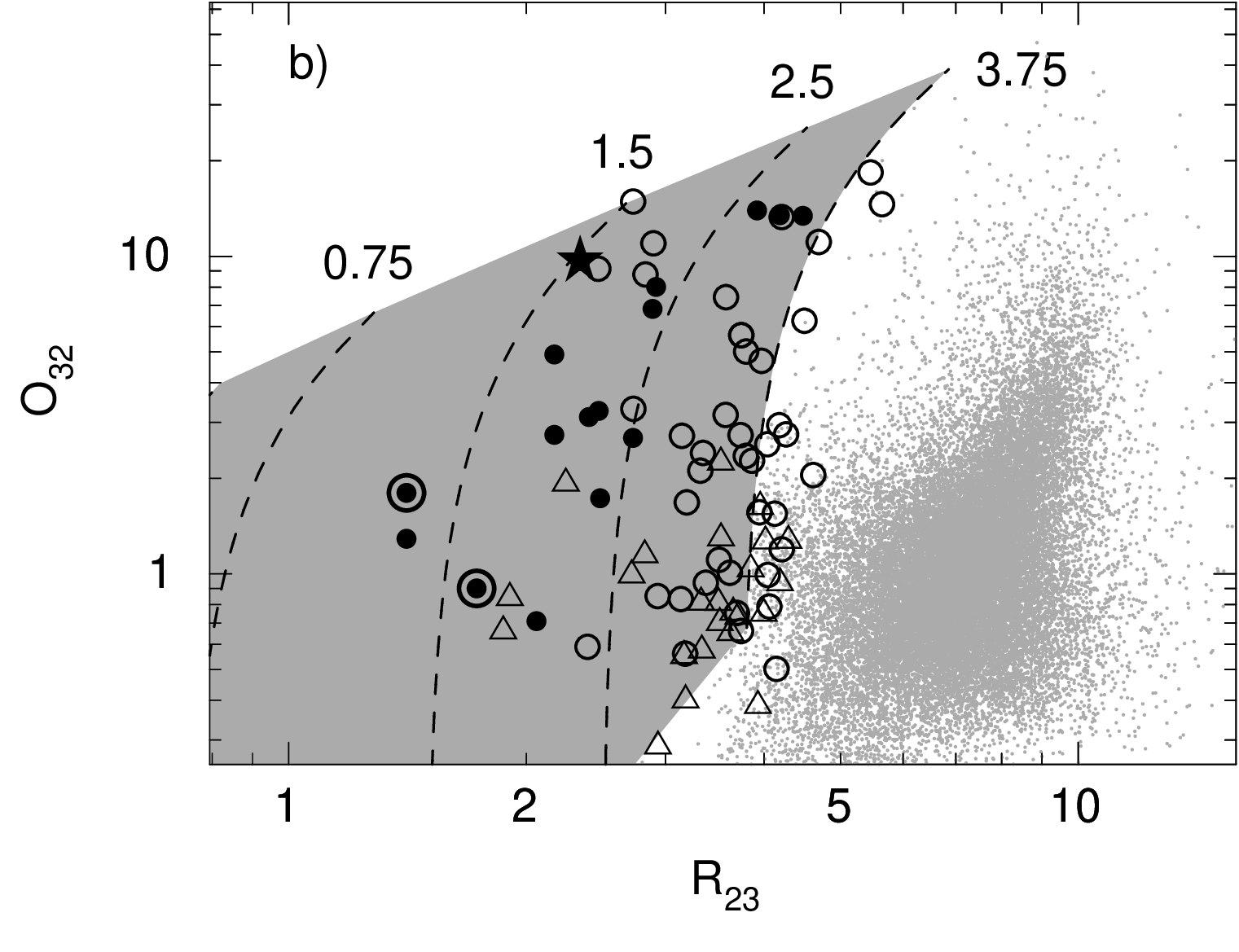}
\caption{{\bf a)} Diagnostic diagram [O~{\sc iii}] $\lambda$5007/H$\beta$ - 
[O~{\sc ii}] $\lambda$3727/H$\beta$. Open circles and open triangles are the 
extremely low-metallicity SFG candidates selected in this paper with oxygen
abundances derived by the direct and semi-empirical methods, respectively.  
Other symbols are the same as in
Fig. \ref{fig1}. The shaded region indicates the location of the extremely
low-metallicity SFGs with 12 + logO/H in the range 6.62 -- 7.39. Dotted 
and dashed lines are the lines of constant O$_{32}$ and of constant 
R$_{23}$ -- 0.08O$_{32}$ (or constant 12 + log O/H), respectively, labelled with their values. The values
R$_{23}$ -- 0.08O$_{32}$ = 0.75, 1.5, 2.5, and 3.75 correspond to oxygen 
abundances 12~+~log~O/H = 6.69, 6.97, 7.18, and 7.35, respectively.
{\bf b)} The diagram O$_{32}$ -- R$_{23}$. Symbols, lines, labels 
and shaded region follow the definitions provided for {\bf a)}.}
\label{fig6}
\end{figure*}

The lowest-metallicity segment of the log(R$_{23}$--0.08$\times$O$_{32}$) -- 
12~+~logO/H relation with 66 selected extremely low-metallicity SDSS SFGs (open 
circles and open triangles)
and some well-studied galaxies from the literature (filled symbols) is shown
in Fig. \ref{fig4}. Some well-known, extremely low-metallicity SFGs are labelled in the figure.
The oxygen abundances for all galaxies shown in the Figure are derived by 1) the direct method if the [O~{\sc iii}]~$\lambda$4363~$\AA$ emission line is 
detected (open circles) or by the 2) semi-empirical method otherwise (open
triangles) while the relation for
the strong-line method determined by Eq. \ref{eq:fit} is shown by the solid 
line. Thirty six SDSS galaxies shown in 
Table \ref{tab2} were previously recovered by our team 
\citep[e.g. ][]{G15,G17,IT07,I12,I18}. Three objects, I Zw 18SE, I Zw 18NW, and 
J0811$+$4730, are presented in both the SDSS and comparison samples and 
connected by solid red lines.

It is seen in Fig. \ref{fig4} that selected SDSS galaxies (open circles
and open triangles) are evenly distributed around the solid line implying that 
our newly proposed strong-line method
(Eq. \ref{eq:fit}) reasonably well reproduces galaxy metallicities. A considerable number of SFGs have 12~+~logO/H lower than in the 
prototypical galaxy I~Zw~18. Two of these galaxies, J0811$+$4730 
(see also Sect. \ref{sec:method}) and
J1234$+$3901 are very close to or below
the luminosity-weighted value 12~+~logO/H = 7.0, representing the 
lowest-metallicity galaxies known so far. In fact, using  
Large Binocular Telescope observations that have high signal-to-noise ratio 
\citet{I18} have confirmed 
that J0811$+$4730 has the lowest luminosity-weighted metallicity among all 
low-redshift SFGs with $z$ $\la$ 0.1 \citep{I18}.
We note that metallicities in regions 3 and 4 of the galaxy
SBS~0335$-$052W are even lower \citep{I09}. However, the
[O~{\sc iii}]~$\lambda$4363~$\AA$ 
emission line is not detected in spectra of these regions and their oxygen
abundances are derived by the semi-empirical method. Furthermore, two other
brighter regions 2 and 1 of SBS~0335$-$052W have higher oxygen abundances 
(Fig.~\ref{fig4}) indicating an abundance gradient in this galaxy. This gradient
results in a higher luminosity-weighted oxygen abundance of 7.12 averaged over 
the entire galaxy SBS~0335$-$052W. \citet{An19} reported the very low oxygen
abundance 12~+~logO/H~=~6.96$\pm$0.09 in the H~{\sc ii} region \#7 of DDO~68
derived by the direct method. However, the oxygen abundance in this region 
is derived with low accuracy and needs to be confirmed with higher precision.
Furthermore, oxygen abundances in other H~{\sc ii} regions of DDO~68 are
considerably higher than in the region \#7 \citep{P05,IT07,IT09}.
We note that spectra of all these regions are not present in the SDSS data base.

One of the lowest-metallicity SFGs among
selected SDSS galaxies is J1234$+$3901. A single high-excitation H~{\sc ii}
region is observed in this galaxy with an oxygen
abundance of 6.98 and 7.03 derived by the strong and direct methods, 
respectively. 
The SDSS spectra of this galaxy and another lowest-metallicity galaxy
J0811$+$4730 from Table \ref{tab2} are shown in Fig. \ref{fig5}. The 
[O~{\sc iii}]~$\lambda$4363~$\AA$ emission line is clearly detected in both 
spectra allowing the oxygen abundance determination by the direct method. 
The galaxy J1234$+$3901 at the redshift of 0.1330 is one of the 
most distant and luminous galaxies in Table \ref{tab2}. Additional observations 
with higher signal-to-noise ratio are needed to confirm the extremely low 
metallicity of this galaxy.

\section{Diagnostic diagrams} \label{diagn}

\citet{I12,I18} and \citet{G17} proposed to use the 
Baldwin, Phillips and Terlevich (BPT) \citep{BPT81} diagnostic diagram
[O~{\sc iii}]$\lambda$5007/H$\beta$ vs. [N~{\sc ii}]$\lambda$6584/H$\alpha$
for the selection of the extremely low-metallicity SFGs. They have shown
that these galaxies are located in the lower-left corner of the diagram far from
the main sequence of SFGs in the region defined by relations
[O~{\sc iii}]$\lambda$5007/H$\beta$ $<$ 4 and 
[N~{\sc ii}]$\lambda$6584/H$\alpha$ $<$ 0.03 corresponding to 
[N~{\sc ii}]$\lambda$6584/H$\beta$ $<$ 0.1. However, the use of this diagram 
is limited because of the weakness of the [N~{\sc ii}]~$\lambda$6584~$\AA$ 
emission line. In many galaxies this line is not detected. 
Alternatively, the diagram 
[O~{\sc iii}]$\lambda$5007/H$\beta$ vs. [O~{\sc ii}]$\lambda$3727/H$\beta$
can be used as the [O~{\sc ii}]~$\lambda$3727~$\AA$
emission line in galaxies with low metallicities is approximately 20 times
brighter than the [N~{\sc ii}]~$\lambda$6584~$\AA$ emission line.

We note that the [O~{\sc ii}]$\lambda$3727/H$\beta$ ratio is sensitive 
to dust extinction. However, in low-metallicity galaxies the internal 
extinction is low. It is derived from the Balmer hydrogen decrement after 
correction of emission line intensities for the Milky Way extinction. Both
extinctions are applied to obtain extinction-corrected line 
intensities in the [O~{\sc iii}]$\lambda$5007/H$\beta$ -- 
[O~{\sc ii}]$\lambda$3727/H$\beta$ diagram.
We show this diagram in Fig. \ref{fig6}a. 
The shaded region indicates the 
location of lowest-metallicity galaxies with 12 + logO/H $\leq$ 7.35. Dashed
lines indicate the lines of equal R$_{23}$ -- 0.08O$_{32}$ corresponding to
equal metallicities according to Eq. \ref{eq:fit}. 
The SFGs selected in this paper and shown in Table \ref{tab2} are 
represented in the Figure by open circles and open triangles and are located in 
the shaded region.
The range of [O~{\sc ii}] $\lambda$3727/H$\beta$ $<$ 2 is somewhat larger than 
$\la$ 1 tested by galaxies with 12 + logO/H derived by the direct method (filled
symbols in Fig. \ref{fig6}a). Therefore, 
metallicities in galaxies with [O~{\sc ii}]~$\lambda$3727/H$\beta$ = 1 -- 2
need to be confirmed with the direct method.

Additionally, for selection of extremely low-metallicity galaxies, the 
O$_{32}$ -- R$_{23}$
diagram can be used, as proposed by \citet{I18}. This diagram for selected SDSS
galaxies (open circles and open triangles) and galaxies from the comparison sample (filled symbols)
is shown in Fig.~\ref{fig6}b. Similar to the diagram in Fig.~\ref{fig6}a these
galaxies are located far from the main-sequence SFGs (grey dots). 
We note a wide range of O$_{32}$ with the highest values reaching values up to
$\sim$ 15. However, most of the extremely low-metallicity SFGs have O$_{32}$ $<$3.
The shaded region corresponds to that in Fig.~\ref{fig6}a. It indicates
the location of extremely low-metallicity SFGs with 12 + logO/H $\leq$ 7.35.

\section{Conclusions \label{S4}}

We present results of a search for the extremely low-metallicity star-forming
galaxies (SFGs) from the Data Release 14 of the Sloan Digital Sky Survey 
(SDSS DR14). 
Our main results are as follows.
 
   1.  A new, simple, purely empirical strong-line method based on the 
[O~{\sc ii}]~$\lambda$3727~$\AA$ and [O~{\sc iii}]~$\lambda$4959~$\AA$, $\lambda$5007~$\AA$ 
emission lines is developed for the oxygen abundance determination at low
metallicities and to search for such extremely low-metallicity galaxies. 
This method is a modification of the well-known
R$_{23}$ method and takes into account the dependence of oxygen emission-line
fluxes not only on the metallicity but also on the ionization parameter
in the H~{\sc ii} region. We adopt that the 
O$_{32}$ = [O~{\sc iii}]$\lambda$5007/[O~{\sc ii}]$\lambda$3727 emission-line 
ratio serves as a measure of the ionization parameter. 
This method was calibrated using observations with high signal-to-noise ratios
(S/N)
of a large sample of SDSS galaxies with detected 
[O~{\sc iii}]~$\lambda$4363~$\AA$
emission applying the direct method in combination with other samples
of SFGs with reliably derived oxygen abundances via the direct method. 

2. Several selection criteria were applied to build a sample of the extremely
low-metallicity SFGs. First, the SDSS galaxies were preselected using
a criterion [N~{\sc ii}]$\lambda$6584/H$\beta$ $\leq$0.2. This allows 
us to exclude high-metallicity objects located on the upper branch of the 
R$_{23}$ -- 12~+~logO/H diagram. Then, all SFGs with 
[O~{\sc iii}]$\lambda$4959/H$\beta$ $\leq$ 0.2 were excluded because 
the [O~{\sc iii}]~$\lambda$4363~$\AA$ emission line was not detected in these 
SFGs and thus their abundances derived by the strong method were
not tested by the direct method. Finally, the remaining SFGs satisfying
condition R$_{23}$ -- 0.08O$_{32}$ $\la$ 4, corresponding to
12~+~logO/H~$<$~7.4, were included in the final list of candidates to the
lowest-metallicity SFGs.

3. We selected 66 emission-line galaxies with 12~+~logO/H~$\leq$~7.35 in the
entire SDSS DR14 using a new strong-line method. Some selected galaxies
have 12~+~logO/H as low as 7.0 and therefore belong to the lowest-metallicity 
SFGs known so far. Results of the metallicity determination by the
new strong-line method are compared with those derived by a semi-empirical
method \citep{IT07}, by a direct method, and by some strong-line methods
from the literature. We find a good agreement
between 12~+~logO/H derived by all three first methods with deviations less than
0.1 dex, on average. The agreement with some methods selected from the
literature is not as good. However, the SDSS spectra of many selected SFGs 
have relatively low S/N. Furthermore, in 61 percent of SFGs the 
[O~{\sc ii}]~$\lambda$3727~$\AA$ emission line is outside the wavelength range of 
SDSS spectra because of their low redshifts $z$ $\la$ 0.02. Therefore, its 
fluxes were estimated from the fluxes of weak 
[O~{\sc ii}]~$\lambda$7320~$\AA$, 7330~$\AA$ or [N~{\sc ii}]~$\lambda$6584~$\AA$
emission lines. We conclude that new observations with higher S/N and including 
[O~{\sc ii}]~$\lambda$3727~$\AA$ emission 
line are needed to confirm the low metallicity of these selected galaxies.

4. We propose the use of two diagnostic diagrams, 
[O~{\sc iii}]$\lambda$5007/H$\beta$ --
[O~{\sc ii}]$\lambda$3727/H$\beta$ and O$_{32}$ -- (R$_{23}$ -- 0.08O$_{32}$)
for the selection of the extremely low-metallicity SFGs.

%

\acknowledgements

 Y.I.I. and N.G.G. thank the hospitality of the Max-Planck 
Institute for Radioastronomy, Bonn, Germany.    
They acknowledge support from the 
Program of Fundamental
Research of the Department of Physics and Astronomy (Project
No. 0117U000240) of the National Academy of Sciences of Ukraine.
Funding for the Sloan Digital Sky Survey (SDSS) has been provided by the 
Alfred P. Sloan Foundation, the Participating Institutions, the National 
Aeronautics and Space Administration, the National Science Foundation, 
the U.S. Department of Energy, the Japanese Monbukagakusho, and the Max Planck 
Society. The SDSS Web site is http://www.sdss.org/.
The SDSS is managed by the Astrophysical Research Consortium (ARC) for the 
Participating Institutions. The Participating Institutions are The University 
of Chicago, Fermilab, the Institute for Advanced Study, the Japan Participation 
Group, The Johns Hopkins University, Los Alamos National Laboratory, the 
Max-Planck-Institute for Astronomy (MPIA), the Max-Planck-Institute for 
Astrophysics (MPA), New Mexico State University, University of Pittsburgh, 
Princeton University, the United States Naval Observatory, and the University 
of Washington.
 


\begin{appendix}
\section{Selected extremely low-metallicity galaxies}
\renewcommand{\baselinestretch}{1.0}
\begin{table*}
\caption{List of the extremely low-metallicity SDSS galaxies from
the DR14 \label{tab1}}
\begin{tabular}{lcccc|lcccc} \\ 
\hline
Name&\multicolumn{2}{c}{Coordinates (J2000.0)}&  $z$$^{\rm a}$  & $g$$^{\rm b}$  & Name&\multicolumn{2}{c}{Coordinates (J2000.0)}&  $z$$^{\rm a}$  & $g$$^{\rm b}$ \\ \cline{2-3} \cline{7-8}
 &RA&Dec     & & & &RA&Dec & &     \\
\hline
J0006$+$2413          &00:06:48.96&$+$24:13:06.94&0.03418&21.66& J1034$+$1546          &10:34:05.40&$+$15:46:50.14&0.00410&17.64 \\
J0015$+$0104          &00:15:20.68&$+$01:04:36.99&0.00686&19.91& J1036$+$2036          &10:36:39.47&$+$20:36:15.80&0.05487&21.60 \\
J0029$-$0025          &00:29:49.50&$-$00:25:39.89&0.01412&20.23& J1053$+$4713          &10:53:21.33&$+$47:13:20.85&0.06796&21.84 \\
J0042$+$3247          &00:42:33.37&$+$32:47:21.01&0.14264&22.13& J1109$+$2007          &11:09:09.53&$+$20:07:29.75&0.00376&17.23 \\
J0106$+$2345          &01:06:09.17&$+$23:45:33.66&0.05206&21.93& J1119$+$0935          &11:19:28.09&$+$09:35:44.28&0.00360&16.43 \\
J0107$+$0103          &01:07:46.56&$+$01:03:52.06&0.00220&19.93& J1121$+$3744          &11:21:46.68&$+$37:44:21.18&0.00644&17.85 \\
J0113$+$0052          &01:13:40.44&$+$00:52:39.15&0.00381&21.18& J1139$+$1917          &11:39:31.78&$+$19:17:24.92&0.01145&17.69 \\
J0122$+$0048          &01:22:41.61&$+$00:48:42.00&0.05731&21.62& J1153$+$3419          &11:53:28.19&$+$34:19:21.89&0.00746&18.01 \\
J0137$+$2032          &01:37:40.98&$+$20:32:46.50&0.03480&22.25& J1157$+$1713          &11:57:44.11&$+$17:13:29.19&0.01267&17.81 \\
J0143$+$1958          &01:43:15.15&$+$19:58:06.10&0.00170&21.76& J1157$+$5638          &11:57:54.18&$+$56:38:16.71&0.00150&16.89 \\
J0153$+$0104          &01:53:11.96&$+$01:04:40.10&0.06332&21.52& J1206$+$5007          &12:06:08.53&$+$50:07:21.17&0.05137&21.76 \\
J0207$-$0821          &02:07:24.77&$-$08:21:43.60&0.01267&20.22& J1208$+$3727          &12:08:09.75&$+$37:27:24.65&0.00356&17.82 \\
J0222$-$0935          &02:22:38.55&$-$09:35:35.20&0.11477&21.61& J1220$+$4915          &12:20:51.61&$+$49:15:55.48&0.01226&20.88 \\
J0223$-$0918          &02:23:02.68&$-$09:18:22.40&0.05032&20.99& J1223$+$0727          &12:23:58.20&$+$07:27:01.73&0.00412&17.90 \\
J0247$-$0404          &02:47:12.80&$-$04:04:31.36&0.03495&22.07& J1226$+$0952          &12:26:55.72&$+$09:52:56.27&0.00330&16.93 \\
J0314$-$0108          &03:14:26.11&$-$01:08:46.55&0.02741&17.70& J1228$-$0125          &12:28:45.54&$-$01:25:26.90&0.07281&22.09 \\
J0808$+$3244          &08:08:56.34&$+$32:44:19.20&0.14640&20.60& J1234$+$3901          &12:34:15.70&$+$39:01:16.41&0.13297&21.92 \\
J0811$+$4730          &08:11:52.12&$+$47:30:26.24&0.04452&21.32& J1235$+$2755          &12:35:52.35&$+$27:55:54.22&0.00261&15.92 \\
J0834$+$5905          &08:34:37.19&$+$59:05:35.99&0.00480&19.86& J1244$+$3212          &12:44:11.17&$+$32:12:21.69&0.00220&17.97 \\
J0859$+$3923          &08:59:46.93&$+$39:23:05.64&0.00200&17.06& J1250$+$1728          &12:50:31.28&$+$17:28:15.95&0.00290&16.37 \\
J0906$+$2528          &09:06:00.92&$+$25:28:11.33&0.00923&17.41& J1257$+$3341          &12:57:40.55&$+$33:41:39.24&0.00300&17.43 \\
J0911$+$3135          &09:11:59.42&$+$31:35:35.93&0.00250&17.71& J1258$+$1413          &12:58:40.20&$+$14:13:00.79&0.00070&16.20 \\
J0921$+$4038          &09:21:36.55&$+$40:38:53.84&0.07136&22.04& J1308$+$2002          &13:08:28.41&$+$20:02:01.93&0.00495&17.54 \\
J0934$+$5514A         &09:34:02.03&$+$55:14:27.86&0.00249&16.44& J1314$+$2438          &13:14:59.16&$+$24:38:39.15&0.01265&18.10 \\
J0934$+$5514B         &09:34:02.39&$+$55:14:23.20&0.00270&17.58& J1315$+$1745          &13:15:56.30&$+$17:45:38.03&0.00320&15.95 \\
J0944$+$0936          &09:44:44.59&$+$09:36:49.18&0.00180&20.36& J1414$-$0208          &14:14:54.13&$-$02:08:22.94&0.00512&18.00 \\
J0945$+$3835          &09:45:19.55&$+$38:35:52.90&0.07245&21.79& J1424$+$5200          &14:24:19.49&$+$52:00:38.40&0.04946&22.64 \\
J0949$+$3426          &09:49:35.09&$+$34:26:16.36&0.00500&22.24& J1433$+$1544          &14:33:21.29&$+$15:44:21.60&0.02027&21.48 \\
J0950$+$3127          &09:50:19.49&$+$31:27:22.24&0.00180&17.77& J1444$+$4237          &14:44:12.80&$+$42:37:44.01&0.00211&21.39 \\
J0955$+$6442          &09:55:31.45&$+$64:42:50.06&0.00320&17.92& J1522$+$4201          &15:22:55.55&$+$42:01:58.30&0.00194&17.48 \\
J0956$+$2849          &09:56:46.05&$+$28:49:43.78&0.00160&14.64& J1640$+$2845          &16:40:21.43&$+$28:45:55.92&0.00329&17.00 \\
J0959$+$4626          &09:59:05.76&$+$46:26:50.49&0.00200&17.79& J1703$+$2126          &17:03:11.59&$+$21:26:21.20&0.10171&20.21 \\
J1000$+$3032          &10:00:36.54&$+$30:32:09.78&0.00170&17.77& J2104$-$0035          &21:04:55.31&$-$00:35:22.24&0.00465&18.13 \\
\hline
\end{tabular}

$^{\rm a}$Redshift.

$^{\rm b}$SDSS $g$-band magnitude.

\end{table*}

\begin{table*}
\caption{Emission-line intensities and oxygen abundances of the extremely low-metallicity SDSS galaxies from
the DR14 \label{tab2}}
\begin{tabular}{lcccccrrcccrc} \\ 
\hline
&\multicolumn{5}{c}{Line intensities$^{\rm a}$}&&&\multicolumn{3}{c}{12+logO/H$^{\rm b}$} \\ \cline{2-6} \cline{9-11}
Name&[O~{\sc ii}]$^{\rm c}$&[O~{\sc iii}]&[O~{\sc iii}]&[O~{\sc ii}] &[N~{\sc ii}] &$F$(H$\beta$)$^{\rm d}$&EW(H$\beta$)$^{\rm e}$&d&se&s&\multicolumn{1}{c}{$M_g$$^{\rm f}$}&log $M_\star$$^{\rm g}$ \\ 
 &  3727      &   4959      &   4363   &   7325  &   6584    \\
\hline
J0006$+$2413         & 0.439 &0.826&0.121& ... & ... &3.9& 39.5&7.25&7.31&7.30&$-$15.0&7.6 \\ 
J0015$+$0104$^{\rm h}$ &(0.898)&0.252& ... & ... &0.045&5.0&177.1&... &7.03&7.06&$-$15.4&4.2 \\ 
J0029$-$0025         &(0.959)&0.726&0.146& ... &0.048&1.8& 35.1&7.25&7.37&7.34&$-$14.4&6.3 \\ 
J0042$+$3247$^{\rm h}$& 0.804 &0.733&0.100& ... & ... &2.4& 68.0&7.22&7.34&7.32&$-$17.2&8.0 \\ 
J0106$+$2345         & 0.223 &0.652&0.063& ... & ... &0.4&277.0&7.17&7.11&7.12&$-$15.2&6.4 \\ 
J0107$+$0103         &(1.682)&0.455& ... &0.056&0.144&18.5& 20.6&... &7.35&7.31&$-$10.3&5.8 \\ 
J0113$+$0052$^{\rm h}$&(0.680)&0.617&0.091& ... &0.034&6.6& 46.3&7.16&7.22&7.25&$-$12.3&5.9 \\ 
J0122$+$0048$^{\rm h}$& 0.548 &0.856&0.100& ... & ... &7.2&156.0&7.22&7.36&7.33&$-$16.1&6.8 \\ 
J0137$+$2032         & 1.293 &0.559& ... & ... & ... &0.4& 26.7&... &7.34&7.31&$-$15.2&6.7 \\ 
J0143$+$1958$^{\rm h}$& 1.597 &0.433& ... & ... & ... &1.4& 20.8&... &7.33&7.29& $-$7.8&4.4 \\ 
J0153$+$0104$^{\rm h}$& 0.214 &1.311&0.134& ... & ... &8.3&229.1&7.39&7.58&7.38&$-$16.2&6.2 \\ 
J0207$-$0821$^{\rm h}$& 0.883 &0.661& ... & ... &0.105&3.7& 29.3&... &7.31&7.30&$-$14.2&6.9 \\ 
J0222$-$0935$^{\rm h}$& 0.915 &0.781&0.103& ... &0.099&1.9& 59.5&7.26&7.39&7.36&$-$17.2&7.9 \\ 
J0223$-$0918$^{\rm h}$& 1.248 &0.677& ... & ... & ... &0.6& 23.1&... &7.40&7.36&$-$16.0&7.5 \\ 
J0247$-$0404         & 0.628 &0.404& ... & ... & ... &3.7& 85.0&... &7.04&7.11&$-$14.1&7.0 \\ 
J0314$-$0108         & 0.494 &0.826&0.084&0.021&0.113&26.2& 31.3&7.26&7.33&7.31&$-$18.9&7.3 \\ 
J0808$+$3244         & 1.819 &0.340&0.036& ... &0.102&8.6& 50.4&7.31&7.32&7.28&$-$18.9&8.9 \\ 
J0811$+$4730$^{\rm h}$& 0.187 &0.570&0.088& ... & ... &15.5&282.0&6.97&7.02&7.03&$-$17.3&5.8 \\ 
J0834$+$5905$^{\rm h}$&(0.986)&0.552&0.059& ... &0.049&8.6& 22.3&7.22&7.26&7.27&$-$15.6&6.7 \\ 
J0859$+$3923$^{\rm h}$&(1.989)&0.438&0.086&0.065&0.091&5.9& 28.3&7.35&7.41&7.34&$-$12.8&5.4 \\ 
J0906$+$2528$^{\rm h}$&(1.864)&0.582& ... & ... &0.093&7.0& 14.4&... &7.46&7.39&$-$16.2&6.7 \\ 
J0911$+$3135$^{\rm h}$&(1.544)&0.519&0.058& ... &0.077&8.4& 16.2&7.32&7.37&7.33&$-$12.9&5.6 \\ 
J0921$+$4038         & 0.276 &1.342&0.181& ... & ... &7.4&300.5&7.32&7.61&7.42&$-$15.9&6.6 \\ 
J0934$+$5514A$^{\rm h}$&(0.185)&0.678&0.072&0.016&0.006&194.9& 57.6&7.14&7.12&7.09&$-$14.1&6.2 \\ 
J0934$+$5514B$^{\rm h}$&(0.505)&0.557&0.045&0.014&0.014&374.6&147.7&7.26&7.12&7.18&$-$13.1&5.1 \\ 
J0944$+$0936         &(1.844)&0.463&0.050&0.060&0.160&7.6& 18.7&7.34&7.39&7.34&$-$10.5&5.3 \\ 
J0945$+$3835$^{\rm h}$& 1.238 &0.845&0.124& ... & ... &3.5& 93.9&7.33&7.50&7.42&$-$16.8&6.9 \\ 
J0949$+$3426         &(1.280)&0.667&0.039&0.042&0.118&12.8& 33.4&7.55&7.40&7.36&$-$16.2&6.0 \\ 
J0950$+$3127$^{\rm h}$&(2.485)&0.416&0.071& ... &0.124&2.8& 12.4&7.42&7.49&7.39&$-$12.7&5.3 \\ 
J0955$+$6442$^{\rm h}$&(1.339)&0.263&0.047& ... &0.067&6.5& 62.9&7.16&7.17&7.16&$-$12.9&5.1 \\ 
J0956$+$2849$^{\rm h}$&(2.124)&0.203& ... &0.070&0.071&7.6& 17.8&... &7.31&7.25&$-$14.6&4.9 \\ 
J0959$+$4626$^{\rm h}$&(1.486)&0.413&0.050& ... &0.074&9.5& 54.1&7.26&7.29&7.27&$-$12.2&5.2 \\ 
J1000$+$3032$^{\rm h}$&(1.939)&0.422& ... & ... &0.097&3.3& 16.4&... &7.39&7.33&$-$12.5&5.5 \\ 
J1034$+$1546$^{\rm h}$&(1.979)&0.521&0.057& ... &0.099&9.6& 19.8&7.37&7.45&7.38&$-$13.9&6.0 \\ 
J1036$+$2036$^{\rm h}$& 0.917 &0.719&0.076& ... & ... &2.8& 83.3&7.31&7.35&7.33&$-$15.6&7.0 \\ 
J1053$+$4713         & 0.481 &1.005&0.130& ... & ... &5.2& 92.2&7.24&7.44&7.38&$-$16.9&7.3 \\ 
J1109$+$2007$^{\rm h}$&(0.685)&0.723&0.093&0.023&0.056&15.8& 61.5&7.19&7.30&7.30&$-$14.1&6.1 \\ 
J1119$+$0935$^{\rm h}$&(0.870)&0.613&0.058&0.029&0.069&23.2& 29.8&7.27&7.27&7.28&$-$14.0&5.9 \\ 
J1121$+$3744$^{\rm h}$&(1.825)&0.336& ... & ... &0.091&2.0& 12.9&... &7.32&7.28&$-$15.2&6.4 \\ 
J1139$+$1917         &(1.821)&0.459& ... & ... &0.091&7.1& 39.3&... &7.39&7.33&$-$17.1&7.3 \\ 
J1153$+$3419         &(1.876)&0.459& ... &0.062&0.183&4.4&  9.0&... &7.40&7.34&$-$15.4&6.9 \\ 
J1157$+$1713         &(0.994)&0.219& ... & ... &0.050&2.1&  9.9&... &7.05&7.05&$-$17.1&7.2 \\ 
J1157$+$5638         &(0.297)&1.099&0.076&0.010&0.030&52.2& 85.1&7.52&7.47&7.36&$-$13.3&5.2 \\ 
J1206$+$5007$^{\rm i}$ & 0.439 &0.826&0.121& ... & ... &7.4&218.5&7.19&7.31&7.30&$-$15.5&6.0 \\ 
J1206$+$5007$^{\rm i}$ & 0.328 &0.813&0.085& ... & ... &7.5&195.4&7.20&7.28&7.26&$-$15.3&6.7 \\ 
J1208$+$3727$^{\rm h}$&(1.889)&0.362& ... &0.062&0.083&5.3& 18.7&... &7.35&7.30&$-$14.6&6.2 \\ 
J1220$+$4915$^{\rm h}$& 0.225 &0.997&0.128& ... & ... &13.9&169.2&7.20&7.38&7.28&$-$13.0&5.6 \\ 
J1223$+$0727         &(1.347)&0.695&0.065& ... &0.067&8.7& 26.3&7.36&7.43&7.38&$-$13.9&5.8 \\ 
J1226$+$0952         &(1.374)&0.390&0.062&0.045&0.102&6.5& 35.3&7.21&7.25&7.24&$-$14.2&5.4 \\ 
J1228$-$0125$^{\rm h}$& 1.616 &0.558& ... & ... &0.043&1.8&107.0&... &7.40&7.35&$-$16.4&6.9 \\ 
J1234$+$3901         & 0.131 &0.650&0.069& ... & ... &12.4&242.0&7.03&7.07&6.98&$-$17.6&7.4 \\ 
J1235$+$2755$^{\rm h}$&(1.493)&0.631& ... &0.049&0.118&12.2& 31.3&... &7.42&7.37& $-$7.9&5.6 \\ 
J1244$+$3212$^{\rm h}$&(0.796)&0.638&0.038&0.026&0.086&23.9&  7.2&7.50&7.27&7.28&$-$15.7&6.3 \\ 
J1250$+$1728         &(2.647)&0.194& ... & ... &0.132&3.7& 15.2&... &7.41&7.31&$-$14.6&5.8 \\ 
J1257$+$3341$^{\rm h}$&(1.741)&0.576&0.039&0.057&0.040&29.9&108.8&7.52&7.44&7.37&$-$13.8&5.4 \\ 
J1258$+$1413$^{\rm h}$&(1.500)&0.469&0.068&0.049&0.073&19.0& 16.3&7.27&7.33&7.30&$-$12.5&5.8 \\ 
\hline
\end{tabular}
\end{table*}

\setcounter{table}{1}

\begin{table*}
\caption{{\sl continued.} Emission-line intensities and oxygen abundances of the extremely low-metallicity SDSS galaxies 
from the DR14}
\begin{tabular}{lcccccrrcccrc} \\ 
\hline
&\multicolumn{5}{c}{Line intensities$^{\rm a}$}&&&\multicolumn{3}{c}{12+logO/H$^{\rm b}$} \\ \cline{2-6} \cline{9-11}
Name&[O~{\sc ii}]$^{\rm c}$&[O~{\sc iii}]&[O~{\sc iii}]&[O~{\sc ii}] &[N~{\sc ii}] &$F$(H$\beta$)$^{\rm d}$&EW(H$\beta$)$^{\rm e}$&d&se&s&\multicolumn{1}{c}{$M_g$$^{\rm f}$}&log $M_\star$$^{\rm g}$ \\ 
 &  3727      &   4959      &   4363   &   7325  &   6584    \\
\hline
J1308$+$2002$^{\rm h}$&(1.592)&0.676& ... & ... &0.080&3.6&  7.3&... &7.47&7.40&$-$15.5&6.3 \\ 
J1314$+$2438         &(2.074)&0.277& ... & ... &0.104&1.9&  6.2&... &7.34&7.28&$-$16.6&6.9 \\ 
J1315$+$1745         &(1.624)&0.647&0.070&0.053&0.145&23.3& 14.8&7.38&7.46&7.39&$-$15.6&6.4 \\ 
J1414$-$0208$^{\rm h}$&(1.417)&0.523&0.051&0.047&0.025&11.4& 54.9&7.30&7.34&7.31&$-$14.2&6.1 \\ 
J1424$+$5200         & 1.117 &0.427& ... & ... &0.065&0.2& 12.5&... &7.21&7.22&$-$15.9&7.1 \\ 
J1433$+$1544         & 0.914 &0.838&0.078&0.040&0.066&2.0& 34.4&7.37&7.45&7.39&$-$13.5&7.1 \\ 
J1444$+$4237$^{\rm h}$&(0.850)&0.833&0.100&0.028&0.049&17.9& 30.2&7.27&7.41&7.37&$-$14.4&5.4 \\ 
J1522$+$4201         &(2.598)&0.333& ... & ... &0.130&4.0& 18.5&... &7.47&7.37&$-$12.3&5.5 \\ 
J1640$+$2845         &(1.998)&0.500& ... & ... &0.100&3.2& 32.3&... &7.44&7.37&$-$14.7&5.8 \\ 
J1703$+$2126         & 1.816 &0.425& ... & ... &0.077&0.7&  8.9&... &7.37&7.32&$-$19.0&8.6 \\ 
J2104$-$0035$^{\rm h}$&(1.171)&0.387& ... & ... &0.042&9.6& 20.7&... &7.20&7.20&$-$14.9&6.2 \\ 
\hline
\end{tabular}

$^{\rm a}$Extinction-corrected line intensity relative to the H$\beta$ emission line intensity. Values in parentheses are calculated
[O~{\sc ii}] $\lambda$3727 line intensities in galaxies where this line is outside the wavelength range of the SDSS spectrum.

$^{\rm b}$s - strong line method, se - semi-empirical method, d - direct method.

$^{\rm c}$The value in parentheses is derived from the [O~{\sc ii}]7325 flux if available or from the [N~{\sc ii}]6584 flux otherwise.

$^{\rm d}$Observed flux of the H$\beta$ emission line in 10$^{-16}$erg s$^{-1}$ cm$^{-2}$.

$^{\rm e}$Equivalent width of the H$\beta$ emission line in $\AA$.

$^{\rm f}$Absolute SDSS $g$ magnitude.

$^{\rm g}$log of stellar mass in log M$_\odot$.

$^{\rm h}$Discussed in \citet{IT07}, \citet{P08}, \citet{I06b,I12,I18}, \citet{G15,G17}.


$^{\rm i}$Different observations of the same galaxy.

\end{table*}

\renewcommand{\baselinestretch}{1.5}
\end{appendix}

\end{document}